\documentstyle[aps,preprint,epsbox]{revtex}
\def\tightenlines{\def\baselinestretch{\@singleleading}}

\begin{document}
\draft
\preprint{}
\title{A Theory of Ferroelectric Phase Transition in SrTiO$_3$ 
induced by Isotope Replacement}

\author{
Yasusada Yamada$^{1)}$, Norikazu Todoroki$^{2)}$ and Seiji Miyashita$^{2)}$
}
\address{{\rm 1)}Advanced Research Institute for Science and Engineering, 
Waseda University
3-4-1, Okubo, Shinjuku-ku, Tokyo 169-0072, Japan\\
{\rm 2)}Department of Applied Physics, The Tokyo University, Bunkyo-ku, Hongo 7-3-1, Tokyo 113-8656\\
}

\date{\today}
\maketitle
\begin{abstract}
A theory to describe the dielectric anomalies and 
the ferroelectric phase transition induced 
by oxygen isotope replacement 
in SrTiO$_3$ is developed. 
The proposed model gives consistent explanation between 
apparently contradictory experimental results on macroscopic 
dielectric measurements versus microscopic lattice dynamical 
measurements by neutron scattering studies. 
The essential feature is 
described by a 3-state quantum order-disorder system characterizing 
the degenerated excited states in addition to the ground state of TiO$_6$ cluster. 
The effect of isotope replacement is taken into account 
through the tunneling frequency between the excited states. 
The dielectric properties are analyzed by the mean field approximation (MFA), 
which gives qualitative agreements with experimental 
results throughout full range of the isotope concentration.
The phase diagram in the temperature-tunneling frequency
coordinate is studied by a QMC method 
to confirm the qualitative validity of the MFA analysis.
\end{abstract}

\section{Introduction}
Recently, the giant anomalous dielectricity and induced 
ferroelectricity by isotope replacement of $^{18}$O for $^{16}$O 
in SrTiO$_3$ have attracted much attention from the view point of 
quantum ferroelectricity.\cite{1,2,3,4} 
In 1999, Ito et al.\cite{1} first reported the appearance of 
broad maximum of the dielectricity at 
$T=23{\rm K}$ in SrTi$^{18}$O$_3$ suggesting the 
stabilization of ferroelectric phase upon isotope 
replacement of $^{16}$O to $^{18}$O. Later, 
the experiments have been refined by using 
single domain specimens and also using partially 
doped SrTi($^{16}$O$_{1-x}$$^{18}$O$_{x}$)$_3$ 
(hereafter STO18-$x$) materials.

\begin{figure}
\epsfile{file=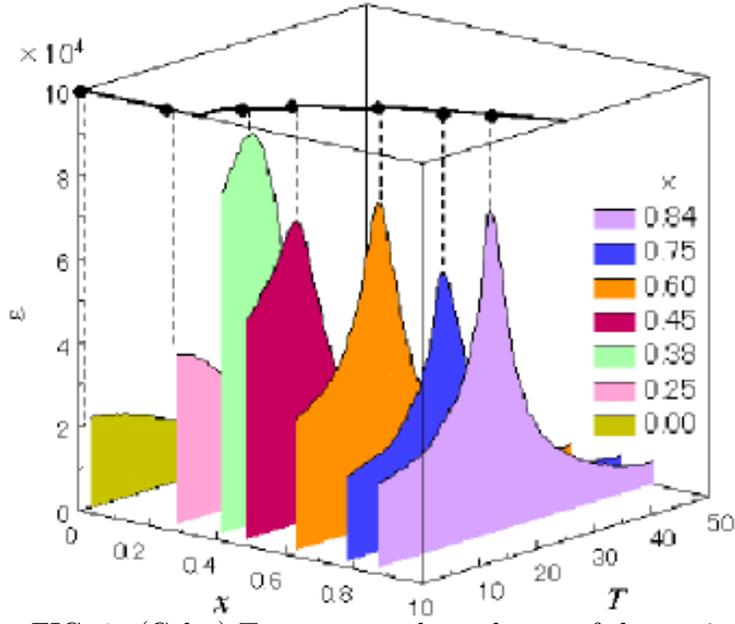,width=10cm}
\caption[]{(Color) Temperature dependences of the static dielectric constant of STO18-$x$ at various $x$-values observed 
by Wang and Ito.\cite{3} The thick solid line on the $T$-$x$ plane (top of the graph) indicates the phase boundary between the paraelectric phase and the ferroelectric phase.\label{fig1}}
\end{figure}

The results by Wang and Ito\cite{3} 
on the dielectric constants are reproduced in Fig. \ref{fig1}
 in somewhat modified form using a three dimensional 
representation. 
As is clearly observed in the figure, 
the peak of static dielectric constant at 
$T=0$ increases as the $^{18}$O concentration $x$ is 
increased to $x_{\rm C}\simeq 33\%$. Upon further increasing $x$, 
the peak position $T_{\rm C}(x)$ shifts continuously,
 and reaches $T_{\rm C}=21.4{\rm K}$ 
at $x=84\%$. When projected on $x-T$ plane, the curve connecting 
the peak positions (the thick solid line in the figure) effectively
 represents the para-ferroelectric phase diagram.

Ito et al. analyzed the experimental data of STO18-$x$
 based on the 'multi component vector model'
 by Schneider et al.,\cite{5,6,6b} 
which deals with the anharmonic lattice vibrational system 
with emphasis on the quantum effect manifested in zero-point vibration. 
In their analysis, the observed $x-T$ phase diagram is reproduced if the 'critical quantum displacive limit' (see Sec. 2) is assumed to be realized at $x=x_{\rm C}  (=33\%)$.
The temperature dependences of the dielectric constant $\varepsilon (T;x)$ with various values of $x$ are analyzed based on Barrett formula\cite{7,8}:
\begin{eqnarray}
\label{eq1-1}
\varepsilon (T;x)=\frac{C}{\frac{T_1}{2}\coth\frac{T_1}{2T}-T_0}
\end{eqnarray}
in the temperature region of $30{\rm K}<T<100{\rm K}$, taking $T_1$ and $T_0$ as adjustable parameters,
and $C$ is a constant.
It is concluded that when $T_1$ is varied in the range of $71{\rm K}<T_1<87{\rm K}$ while $T_0$ kept constant, the results show satisfactory agreements throughout the observed $x$-values. Thus, as far as the macroscopic dielectric characteristics such as phase diagram and static dielectric constant are concerned, the experimental results seem to be explained satisfactorily by the anharmonic lattice vibrational system with quantum effect taken into account explicitly.

\begin{figure}
\epsfile{file=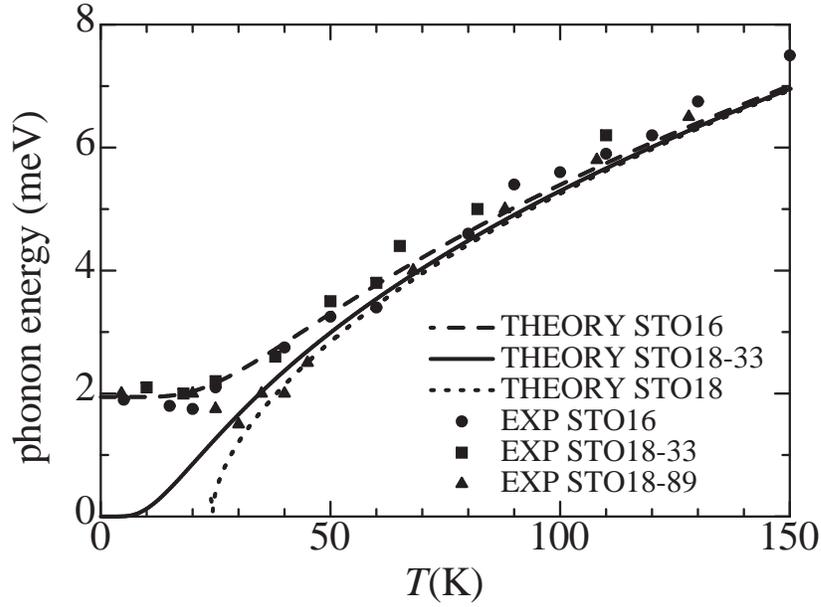,width=12cm}
\caption[]{Comparison of the experimental soft TO-mode frequencies $\hat{\omega}_0(T)$ 
observed by neutron inelastic scattering and the theoretical curves based 
on the anharmonic quantum lattice vibrational system for the cases of $x=0\%$(dashed carve), at the quantum displacive limit $x=x_{\rm C}$(solid curve), and $x=100\%$(dotted carve).\label{fig2}}
\end{figure}

     Recently, however, microscopic observations of the soft TO phonon have been carried out on STO18-$x$ using neutron inelastic scattering technique.\cite{9,10} The results show essential discrepancies with the expected characteristics of the TO phonon frequency $\omega_0(T)$. The results are summarized in Fig. \ref{fig2}.
 In the figure the calculated curves of $\omega_0(T)$ for an anharmonic phonon system (see Sec. 2) 
are plotted for the cases of $x=0\%$, $x=x_{\rm C}$ and $x=100\%$, 
together with the observed points at $x=0\%, 33\%$, and $89\%$. 
As is clearly seen in the figure, the observed values for $x=33\%$ and $89\%$ are in complete disagreement with the calculated curves. 
That is, while the calculated curves give the remarkable characteristic that they reach the stability limit ($\omega=0$) at $T\ge 0$, the observed points do not seem to exhibit such tendencies. 
In fact, the soft TO mode frequency is practically $x$-independent. These results seem to indicate that there should be some other freedom of motion which is responsible 
for the anomalous dielectric behavior of STO18-$x$.
It is the purpose of the present investigation to present a possible model to give a consistent explanation for the apparent discrepancy between the macroscopic versus the microscopic observations.
We propose a model where we introduce a variable specified by 'bipolarons'
due to the $^{18}$O substitution. 
The model is a generalized Blume-Capel (BC) model \cite{BC} with quantum fluctuation. 
In this model, the new degree of freedom causes the order-disorder phase transition and also 
zero temperature fluctuation of polarization.
We analyse this model by the mean field approximation (MFA) and Quantum Monte Carlo (QMC) simulation,
and compare the results with the observed data.
\section{Quantum displacive system vs quantum order-disorder system}

     Ferroelectric phase transition is conventionally divided into two categories: 
displacive type and order-disorder type. 
Correspondingly, the quantum effect of polarization fluctuation is described somewhat differently.
 In this section we briefly review the quantum effect in both cases.
\subsection{Quantum displacive system}
     The Hamiltonian to describe the system is given as follows\cite{11}:
\begin{eqnarray}
\label{eq2-1}
{\cal H}=
 \frac{1}{2}\sum_i(P_i^2+\omega_{\rm loc}^2Q_i^2)
+\frac{\lambda}{4}\sum_iQ_i^4
+\sum_{i,j} V_{ij}Q_iQ_j,
\end{eqnarray}
where $P_i$ and $Q_i$ are respectively the momentum and amplitude of the local optical vibrational mode at the  $i$-th site, $\omega_{\rm loc}$ is the characteristic frequency, and $V_{ij}$ is the interaction between the induced polarization on different sites. 
Through the renormalization process of the anharmonic part, the system is represented by a quasi-harmonic system with the renormalized frequency $\hat{\omega_k}$,
\begin{eqnarray}
\label{eq2-2}
{\cal H}=\frac{1}{2}\sum_k(P_k^2+\hat{\omega}_k^2Q_k^2),
\end{eqnarray}
\begin{eqnarray}
\label{eq2-3}
\hat{\omega}_0^2=\omega_0^2+\lambda\langle Q^2\rangle,
\end{eqnarray}
\begin{eqnarray}
\label{eq2-4}
\omega_0^2=\omega_{\rm loc}^2-V(q=0).
\end{eqnarray}
Here, $\omega_0$ stands for the 'bare' 
harmonic phonon frequency of the uniform ($q=0$) TO mode.
Notice that in order to undergo ferroelectric phase transition $\omega_0^2$ should be negative. 
That is, without the anharmonicity the system is intrinsically unstable against uniform TO mode. 
The renormalized frequency $\hat{\omega}_0^2$  is temperature dependent through
\begin{eqnarray}
\label{eq2-5}
\langle Q^2\rangle
=\frac{\hbar}{2\omega_{\rm loc}}\coth\frac{\hbar\omega_{\rm loc}}{2k_{\rm B}T},
\end{eqnarray}
which is extrapolated at higher temperatures as
\begin{eqnarray}
\label{eq2-6}
\langle Q^2\rangle
\rightarrow \frac{k_{\rm B}T}{\omega_{\rm loc}^2}
\end{eqnarray}
giving the 'classical limit'. 
The renormalized frequency $\hat{\omega}_0$ is nothing but the so-called 'soft' mode frequency characterizing the displacive ferroelectricity.

\begin{figure}
\epsfile{file=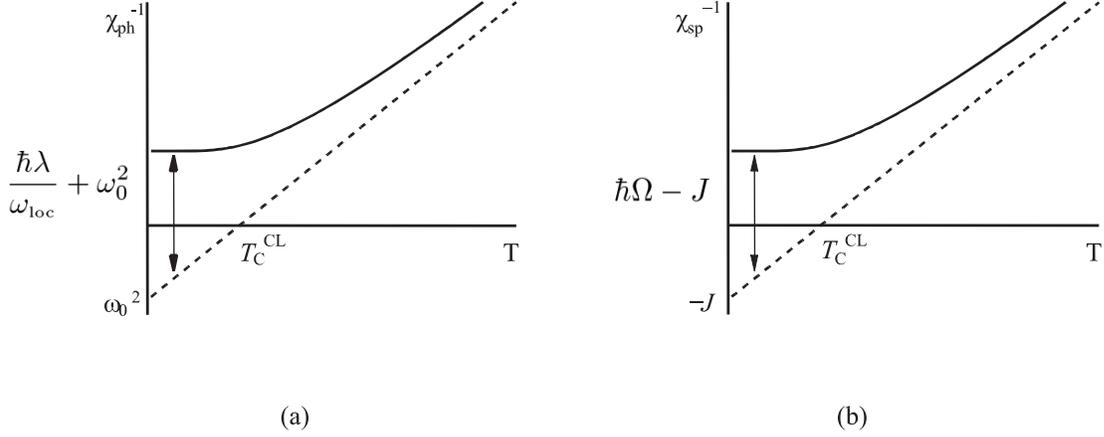,width=15cm}
\caption[]{Schematic representations of the effect of quantum fluctuations appearing in $\chi^{-1}-T$ curves.
(a) The susceptibitity $\chi_{\rm ph}$ (Eq. \ref{eq2-7}) for the case of quantum displacive system where the quantum effect is due to the zero-point vibration, 
and (b) $\chi_{rm sp}$ (Eq. \ref{eq2-9}) for the case of quantum order-disorder system where quantum effect is due to the tunneling of the quasi-spin variable. The dashed lines correspond to the cases for classical limits. The effects of quantum fluctuations are visualized as the differences between the dashed lines and the solid curves.\label{fig3}}
\end{figure}

Since the dielectric susceptibility (or the dielectric constant $\varepsilon$) due to the soft TO mode is given by
\begin{eqnarray}
\label{eq2-7}
\chi_{\rm ph}=\frac{1}{\hat{\omega}_0^2},
\end{eqnarray}
the overall characteristic feature is conveniently overviewed by plotting $\chi_{\rm ph}^{-1}$ vs temperature (Fig. \ref{fig3} (a)). 
As is evident in the figure, the effect of the quantum fluctuation shifts the $\chi_{\rm ph}^{-1}$ from 
the dashed curve representing the classical limit to  the solid curve.

\subsection{Quantum order-disorder system}
In order-disorder type systems, the local potential for the order parameter has the multi-minimum structure. Hence, the variable associated with the polarization is considered to take several discrete values which are effectively expressed by a 'spin' variable. \cite{12,13}

In the simplest case of the double minimum potential, the effective Hamiltonian is expressed in terms of the  $2\times 2$ Pauli spin operators as follows,
\begin{eqnarray}
\label{eq2-8}
{\cal H}=\hbar\Omega\sum_i\sigma_i^x-\sum_{i,j}J_{ij}\sigma_i^z\sigma_j^z,
\end{eqnarray}
where $\Omega$ is the tunneling frequency of the hopping motion of the polarization, and $J_{ij}$ is the interaction between the local polarizations at the sites $i$ and $j$.

Within MFA the spin susceptibility is given by
\begin{eqnarray}
\label{eq2-9}
\chi_{\rm sp}=\left (\frac{1}{1-J(0)\chi_{\rm sp}^{\rm s}}\right )\chi_{\rm sp}^{\rm s},
\end{eqnarray}
where
\begin{eqnarray}
\label{eq2-10}
\chi_{\rm sp}^{\rm s}=\frac{1}{\hbar\Omega}\tanh\frac{\hbar\Omega}{k_{\rm B}T}.
\end{eqnarray}
Here, $\chi_{\rm sp}^{\rm s}$ stands for the susceptibility of an isolated tunneling particle (single particle susceptibility), while the term in the bracket in Eq. (\ref{eq2-10}) is the 'enhancement factor' due to cooperativity. 
Notice the classical limit of $\chi_{\rm sp}^{\rm s}$  is given by
\begin{eqnarray}
\label{eq2-11}
\chi_{\rm sp}^{\rm s}\rightarrow \frac{1}{k_{\rm B}T}.
\end{eqnarray}
In order to overview the characteristic feature, we again plot $\chi_{\rm sp}^{-1}$ vs temperature (Fig. \ref{fig3}(b)). 
We obtain a similar feature to the case of displacive type except that the quantum effect is due to the tunneling motion rather than the zero-point vibration of the particle.

\subsection{Barrett formula}
As has been already pointed out by several authors,\cite{14,15,16} in spite of apparently quite different formalism of $\chi_{\rm ph}$ and $\chi_{\rm sp}$ (Eqs. (\ref{eq2-7}) and (\ref{eq2-9})), 
both can be reduced to the common expression called Barrett formula (Eq. (\ref{eq1-1})) except that the characteristic parameters, $T_0$, $T_1$, and $C$, are expressed in terms of different physical quantities as listed in Table \ref{table1}.

This situation implies that as far as the macroscopic characteristics such as static dielectric constant are concerned, 
it is difficult to distinguish which category the system under investigation belongs to. 
In order to distinguish the category, some microscopic measurements are necessary. 
In the case of SrTiO$_3$, neutron inelastic scattering gave the decisive information. 
In 1968, Yamada and Shirane\cite{16} observed the phonon dispersions of STO16 by neutron inelastic scattering technique. 
They found the well-defined TO phonon branch which shows strong softening particularly at $q=0$. 
The observed temperature variation of the soft mode frequency $\omega_0(T)$ was fitted satisfactorily to Barrett formula in a wide temperature range of $20{\rm K}<T<300{\rm K}$. 
The results clearly indicate that STO16 belongs to the typical 'quantum displacive' system. 

\section{3-state quantum order-disorder system}

As a natural extension of STO16 described in the last section, one expects that STO18-$x$ would be understood within the same framework of quantum displacive system. 
However, recent neutron scattering studies on STO18-33\cite{9} and STO18-89\cite{10} seem to provide essentially contradictory results to this anticipation. 
As is summarized in Fig. \ref{fig2}, the observed temperature dependences of the soft TO phonon frequencies are practically $x$-independent. 
On the other hand, the theoretical values based on the quantum displacive system, 
\begin{eqnarray}
\label{eq3-1}
\omega_0=
\left [
\frac{1}{C}
\left (\frac{T_1}{2}\coth\frac{T_1}{2T}-T_0\right )
\right ]^{1/2}
\end{eqnarray}
with $T_0$ and $T_1$ given by fitting to the observed $\varepsilon (T;x)$, 
show substantially different temperature dependences for the typical cases of 
$x=x_{\rm C}$($=33\%$) and $x=100\%$.

\begin{figure}
\epsfile{file=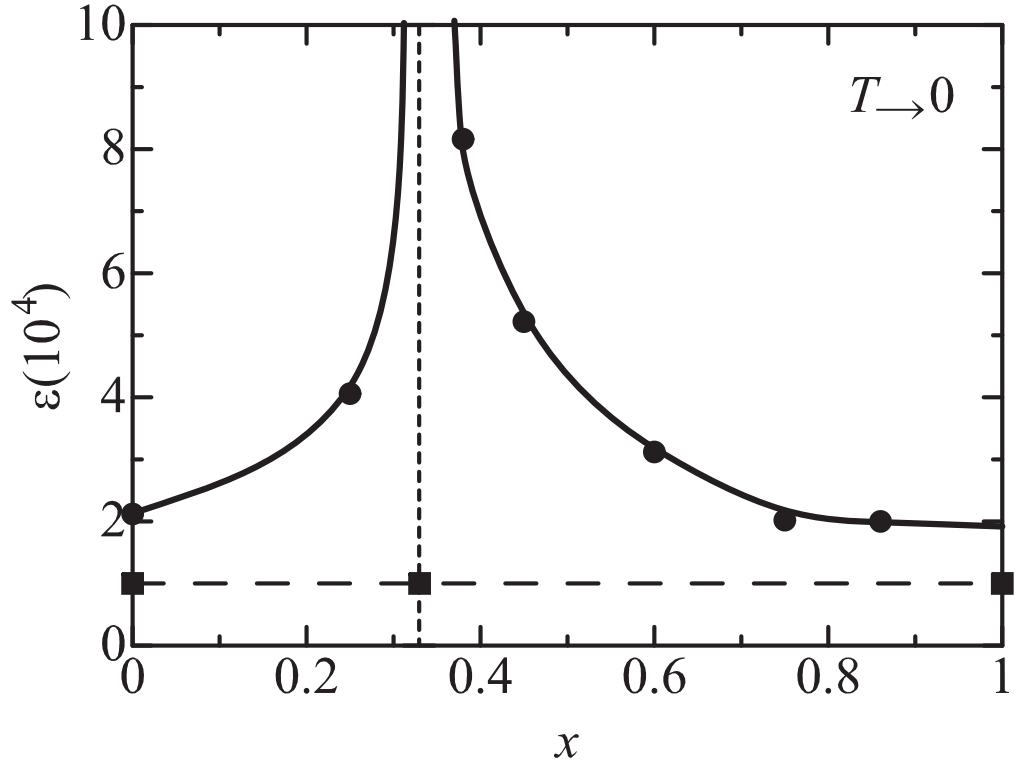,width=15cm}
\caption[]{The observed dielectric constants as extrapolated to $T=0$. 
The solid circles are the values by macroscopic dielectric measurements. 
The solid squares are the values deduced from the microscopic measurements of $\hat{\omega}_0(T)$ using (\ref{eq2-7}). 
The solid and dashed lines are for guides of eyes.\label{fig4}}
\end{figure}

It is instructive to plot at the zero temperature limit ($T\rightarrow 0$) 
as a function of $x$ as given in Fig. \ref{fig4}. 
It is seen that the contribution due to the zero-point motion 
of soft TO phonon which is deduced from the neutron data 
(the solid squares and the dashed line in the figure) is irrespective of the observed critical behavior of $\varepsilon (T\rightarrow 0;x)$ around $x=33\%$. 
This fact suggests that the dielectric behavior of STO18-$x$ is composed of two different components, and that the anomalous part in low temperature region originates from the quantum fluctuations associated with some other freedom of motion than the soft TO phonon mode.

\begin{figure}
\epsfile{file=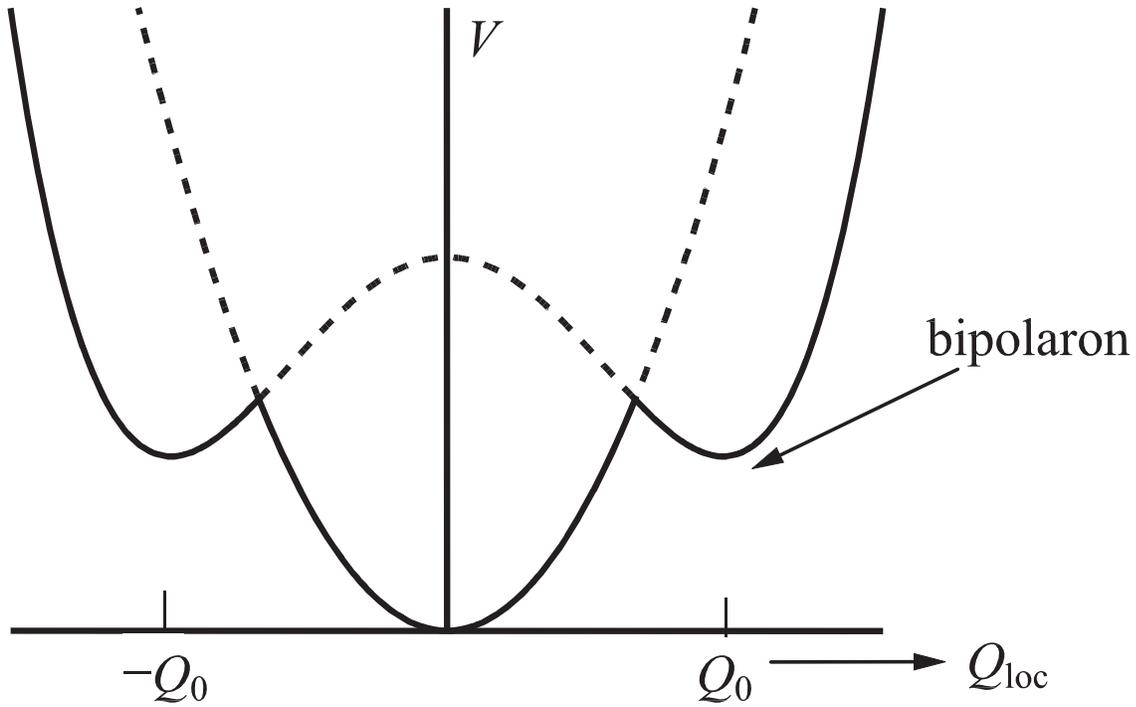,width=15cm}
\caption[]{Schematic diagram of the adiabatic deformation potential of TiO$_6$ cluster as the amplitude of the local polar mode is varied. 
The central valley is associated with the electronic ground state, 
while the side valleys are associated with the first excited state, which is characterized as the bipolaronic state.\label{fig5}}
\end{figure}

Recently, Hasegawa et al.\cite{17} reported the observation of a giant dielectricity induced by laser irradiation of STO16 crystals, 
which suggests the possibility that the anomalous dielectricity in SrTiO$_3$ is due to the polarization fluctuation associated with the electronic excited states. 
The excited state is created by single electron transfer from one of the oxygens to titanium: Ti$^{4+}$O$_{6}^{2-}$$\rightarrow$Ti$^{3+}$O$_{5}^{2-}$O$^{-}$. 
Since the excited state breaks inversion symmetry inducing spontaneous distortion of the octahedron, 
the adiabatic potential for the polar local mode, $Q_{\rm loc}$, would have the characteristic as shown schematically in Fig. \ref{fig5}. 
It is noticed that in addition to the minimum at $Q_{\rm loc}=0$, corresponding to the ground state, there are local minima at $Q_{\rm loc}=\pm Q_0$, corresponding to the excited states. 
The excited state accompanied by the spontaneous distortion may be viewed as 
a pair of an electron polaron around Ti$^{3+}$ and a hole polaron around O$^{-1}$. 
This pair is a kind of 'bipolaronic state'. 

Based on these considerations we construct a model which takes into account the bipolaronic exited states in addition to the ground state of TiO$_6$ cluster. 
If we neglect the small amplitude vibrations around the local minima in Fig. \ref{fig5}, 
the freedom of motion of the system is simply represented by a 'spin' variable which takes three discrete values; $S^z=0, +1$, and $-1$. 
We further assume that 
while the transition between the ground state 
and the excited states $S^z=0\leftrightarrow \pm 1$ is allowed only by thermal excitation, 
quantum hopping (or tunneling) is allowed between the degenerated bipolaronic states 
($S^z=1\leftrightarrow -1$).

The 'spin' part of the freedom of motion is characterized by a 3-state model as the BC model.
In orderer to take into account the quantum hopping, we introduce a quantum transfer term. 
Thus, the Hamiltonian of the model is given by 
\begin{eqnarray}
\label{eq3-2}
{\cal H}
=h\Omega \sum_i\hat{S}_i^x+\frac{1}{2}\sum_{i,j}J_{ij}S_i^zS_j^z+\Delta\sum_i (S_i^z)^2.
\end{eqnarray}
Here, $\hat{S}_i^x$ and $S_i^z$ are the spin operators expressed by $3\times 3$ matrices rather than the $2\times 2$ Pauli spin operators $\sigma$ in Eq. (\ref{eq2-8}):
\begin{eqnarray}
\hat{S}_i^x
=
\left (
\begin{array}{ccc}
0 & 0 & 1 \\
0 & 0 & 0 \\
1 & 0 & 0 
\end{array}
\right ),\hspace{5mm}
S_i^z
=
\left (
\begin{array}{ccc}
1 & 0 & 0 \\
0 & 0 & 0 \\
0 & 0 &-1 
\end{array}
\right ) .
\end{eqnarray}
The interaction $J_{ij}$ denotes the ferroelectric interaction between the polarization.
Here we consider interactions only between the nearest neighbour spins.
It should be noted that $\hat{S}^x$ is not the standard operator for the $x$-component of the angular momentum. 
The parameter $\Delta$ in the third term describes the excitation energy to create the bipolarons. 
Within MFA, the Hamiltonian is given in matrix form by, 
\begin{eqnarray}
\label{eq3-4}
{\cal H}_{\rm MFA}
=
\left (
\begin{array}{ccc}
-Jz\langle S^z\rangle +\Delta & 0 & h\Omega \\
0 & 0 & 0 \\
h\Omega & 0 &Jz\langle S^z\rangle +\Delta 
\end{array}
\right ),
\end{eqnarray}
where $z$ denotes the numbers of the nearest neighbour sites.

Through a standard process, the free energy of the system is explicitly given by 
\begin{eqnarray}
\label{3-5}
F(\xi )&=&-T\ln{\rm Tr}\left ( e^{-{\cal H}_{{\rm MFT}}/T}\right )+\frac{1}{2}Jz\xi^2 \nonumber \\
&=&\frac{1}{2}Jz\xi^2-T\ln\left [ 2\cosh\frac{1}{T}\sqrt{\Omega^2+z^2J^2\xi^2}+\exp\frac{\Delta}{T}\right ],
\end{eqnarray}
where $\xi=\langle S^z\rangle$ stands for the order parameter or the spontaneous polarization.
For conciseness of the expression we have put $k_{\rm B}=\hbar=1$.
The qualitative nature of ordering studied by the MFA are confirmed by QMC simulation as given in the Appendix.
\section{Analysis of experimental results}
     Once the general expression of $F(\xi)$ is given, 
we can obtain various thermodynamical quantities such as dielectric constant $\varepsilon (T;x)$ as a function of the three parameters, $\Omega$, $J$, and $\Delta$ (essentially two independent parameters, $\Omega /J$ and $\Delta /J$). 
In order to explain the experimental results, we have to determine the dependence of the parameters on the concentration $x$. 
We conjecture that among these parameters, the tunneling frequency $\Omega$ would be most sensitive to the isotope replacement. 
Hereafter, we discuss only the limited case that only $\Omega$ is $x$-dependent and $\Delta /J=1$. 
The second assumption implies that the excitation energy to create a bipolaron is comparable to the total nearest neighbour interaction.

\begin{figure}
\epsfile{file=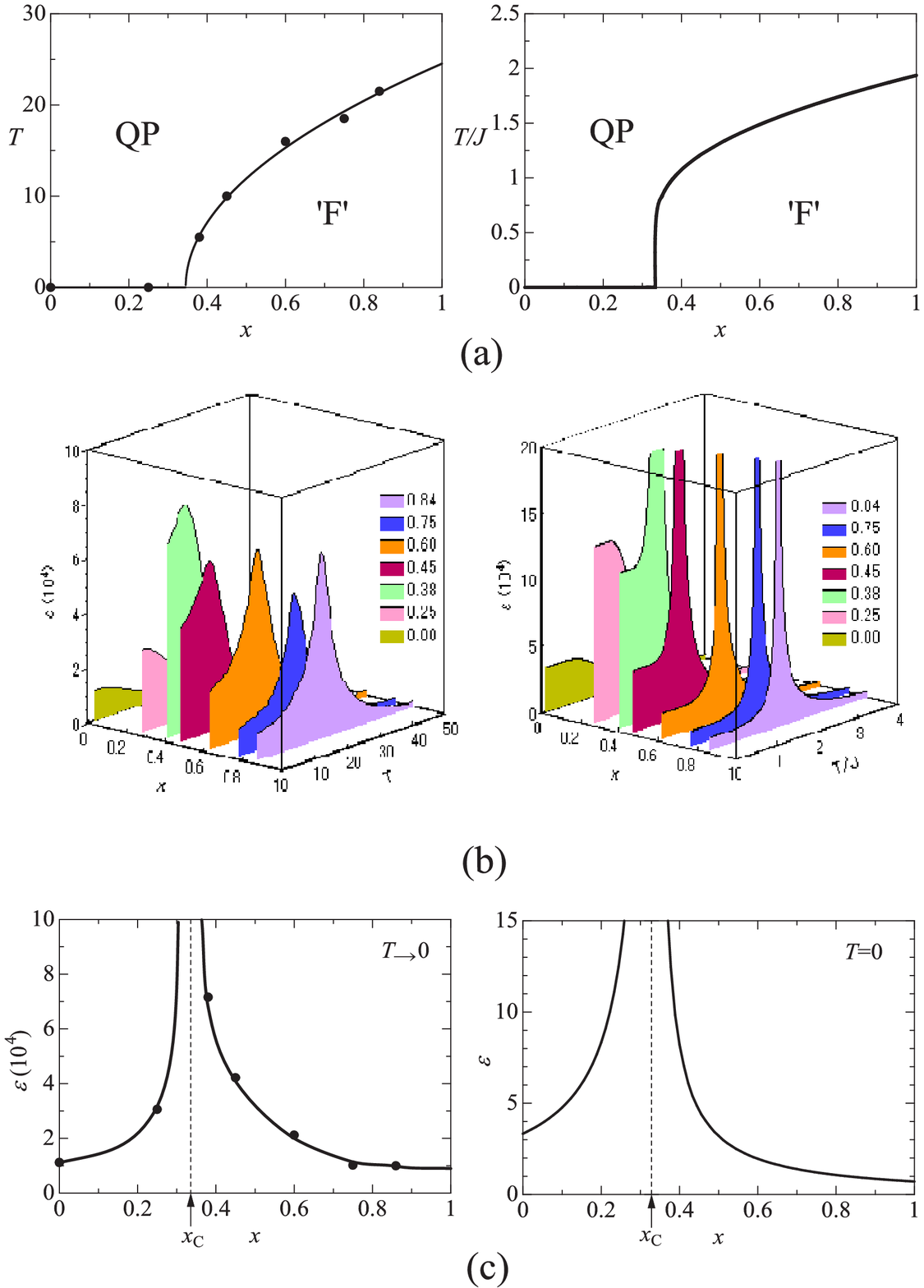,width=14cm}
\caption[]{Comparison of the observed characteristic features in $\varepsilon (T;x)$ (left column) and the corresponding theoretical curves (right column) with parameter values given in the text.
(a)phase diagram
QP: quantum paraelectric phase
'F': extraordinary ferroelectric phase (See the text.)
(b)Temperature and the concentration dependence of $\varepsilon (T;x)$
In the experimental curves the contribution of the TO soft mode fluctuation has been subtracted.
(c)Dependence of $\varepsilon (T\rightarrow 0;x)$ on the concentration.
The experimental values are given by subtracting the contribution due to quantum displacive fluctuations.\label{fig6}}
\end{figure}

It is easily seen that, in the model (\ref{eq3-4}), the quantum mechanical stability limit for the ferroelectric state is
\begin{eqnarray}
\label{eq4-1}
\Omega (x_{\rm C})/J=6.0
\end{eqnarray}
for the case of $z=6$ corresponding to the simple cubic lattice. 
We recognize that this condition is satisfied at $x=33\%$ ($=x_{\rm C}$ ) in STO18-$x$. 
A linear dependence of $\Omega (x)$
\begin{eqnarray}
\label{eq4-2}
\Omega (x)/J=\Omega (x_{\rm C})/J+0.9(x_{\rm C}-x),\hspace{1cm}0\le x\le 1,
\end{eqnarray}
is adopted as the simplest choice. 
The parameter value 0.9 has been chosen so that overall phase diagram becomes 
qualitatively consistent with the observed one (see Fig.\ref{fig6}(a)).
This dependence means that the tunneling frequency changes by $15\%$ within the full range of isotope replacement: $0\%\le x\le 100\%$.

The results of the calculations are summarized in Fig. \ref{fig6}. 
In Fig. \ref{fig6}(a), the ordered phase is specified as 'ferroelectric' within the quotation mark, 
which implies that the ordered state is somewhat different from the ordinary ferroelectric state.
In the case of the ordinary ferroelectric states, the polarization fluctuation is substantially replaced by the spontaneous polarization as $T\rightarrow 0$. 
In contrast, in the present case there remains considerable amount of polarization fluctuation even at $T=0$ which is associated with the tunneling fluctuation of the spins as well as the soft mode vibration. 
In Fig. \ref{fig6} (b) and (c), the 'experimental' values are given after subtracting the contribution due to the quantum displacive part given by Eq. (\ref{eq2-7}) in order to facilitate direct comparison with the calculations based on the 3-state quantum order disorder model. 
Although the calculated results are still far from the situation to carry out quantitative analysis to fit the experimental data, 
the overall qualitative features seem to be satisfactorily reproduced.
\section{Conclusions and discussions}

A theory to describe the dielectric anomalies induced by oxygen isotope replacement in SrTiO$_{3}$ is proposed which gives consistent interpretation between the macroscopic properties (static dielectric constant) and the microscopic properties (TO phonon frequency observed by inelastic neutron scattering). 
The essential feature of the anomaly is characterized by the 3-state quantum order-disorder system (\ref{eq3-2}) associated with the bipolaronic excited states in addition to the ground state of TiO$_6$ cluster. 
By taking into account the effect of the isotope replacement through the frequency of the tunneling between the bipolaronic states, 
the dielectric properties are investigated. 
The results give qualitative agreements with the experimental results throughout full range of the isotope concentration, $0<x<100\%$. 

\begin{figure}
\epsfile{file=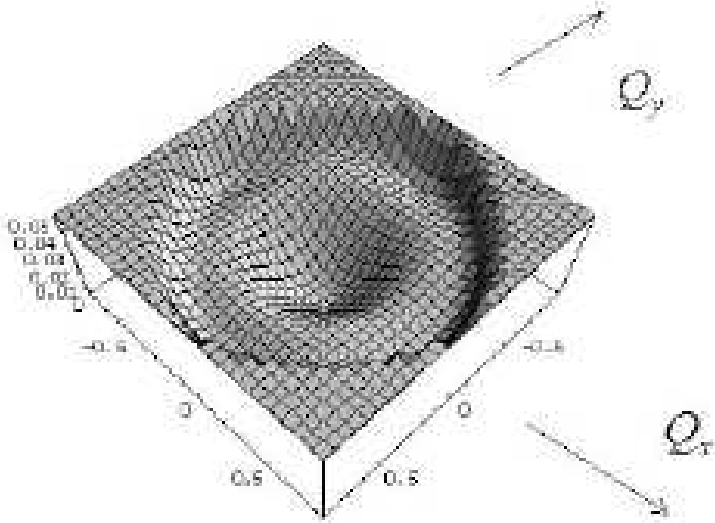,width=10cm}
\caption[]{\label{fig7}The adiabatic potential for the triply degenerated polar modes ($Q_x, Q_y, Q_z$) within ($Q_x,Q_y,0$)-plane. The minima are at ($\pm Q_0,0,0$) and ($0,\pm Q_0,0$), 
and the lowest barriers are at ($\pm, Q_0,\pm Q_0,0$). The tunneling bipolaron is 'rotating' along the ditch as it changes the state.}
\end{figure}

In the present treatment, we have assumed for simplicity that there exists a single (non-degenerate) polar mode, $Q_{\rm loc}$, which couples to the electronic excitation. 
In the case of TiO$_6$, however, the polar modes are triply degenerated forming the basis functions, $Q_x, Q_y, Q_z$, of T$_{1u}$ irreducible representation of the cubic point group m3m.
Therefore, the adiabatic potential of the polar mode should have quasi-spherical symmetry in the ($Q_x, Q_y, Q_z$) space. 
There are at least six minima located on the sphere with a fixed radius $Q_0$ and the lowest potential barriers (saddle points) between the minima through which the bipolaron undergoes tunneling are also situated on the same sphere. 
That is, the 'spin' is rotating along the spherical ditch as it changes the state
(see Fig. \ref{fig7}). 
Correspondingly, the tunneling modes would be split into six levels instead of two as assumed in the present paper. In order to carry out quantitative analysis proper symmetry considerations would be crucially important. 

The direct evidence of the validity of the present picture should be given by the observation of the tunneling excitations using Raman scattering, neutron scattering, etc.. 
Up to this date, any well-defined excitation spectrum has not been reported in addition to the ordinary phonon excitations observed by neutron scattering technique. Recently, Noda et al.\cite{18} observed quasielastic neutron scatterings around the Bragg reflections which exhibit strong temperature dependence. 
Therefore it would be possible that the tunneling modes are heavily damped so that the scattering spectra due to tunneling fluctuation have resulted in so-called 'central peak' spectrum centered at $\omega=0$.  
Similar kind of critical over-damping of the soft mode is commonly observed at various structural phase transitions. 
More detailed spectroscopic observations of the critical fluctuations around $x<x_{\rm C}$  at lower temperatures are needed.

\begin{appendix}
\section{Quantum Monte Carlo simulation}
We study the phase diagram of the model (\ref{eq3-2}) by a QMC method
with the Suzuki-Trotter decomposition. 
In the present model non-diagonal interaction exists only at each site.
Thus, the model is simply transformed to a four-dimensional classical model.
We performed Monte Carlo simulation 
on this model by using a cluster heat bath algorithm.\cite{19}
Recently efficiently of QMC method has been extensively developed by introducing the loop algorithm to
avoid the slowing down due world line algorithm at low temperatures. \cite{20}
The loop algorithm has been also extended to system s with large $S$. \cite{21}
In this study, however, we adopt a cluster heat bath algorithm
which provides an efficient sampling in strongly spacially anisotropic systems.
In the cluster heat algorithm, the Boltzmann weight is calculated straightforwardly
by using one dimensional transfer matrix method. This method is efficient 
for the quasi one-dimensional discrete spin models.
In this method we do not need to provide the graph representation of the cluster which depends on the details of the model.
Thus, this cluster heat algorithm algorithm is very convenient for the present study.

The Suzuki-Trotter decomposition transforms 
the Hamiltonian (\ref{eq3-2}) into
\begin{eqnarray}
{\rm Tr} e^{-\beta{\cal H}}\simeq {\rm Tr}\left ( e^{-\frac{\beta}{N}\left (J\sum_{\langle i,j\rangle}S_i^zS_j^z+\Delta\sum_i(S_i^z)^2\right )}e^{-\frac{\beta\Omega}{N}\sum_{i}\hat{S}_i^x}\right )^N \rightarrow {\rm Tr}e^{-\beta{\cal H}_{\rm eff}}
\end{eqnarray}
where ${\cal H}_{\rm eff}$ is a four-dimensional classical model with a variable $\sigma =-1,0$, and $1$
\begin{eqnarray}
\beta{\cal H}_{\rm eff}=-K_{N} \sum_{i}\sum_{\mu}\sigma_{i,\mu}\sigma_{i+1,\mu}
         -\tilde{J}\sum_{i}\sum_{\langle \mu,\nu\rangle}\sigma_{i,\mu}\sigma_{i,\nu}
         +\tilde{\Delta}\sum_{i}\sum_{\mu}(\sigma_{i,\mu})^2,
\end{eqnarray}
\begin{eqnarray}
K_N&=&\frac{1}{2}\log\left (\cot{\frac{\beta \Omega}{N}}\right ), \\
\tilde{J}&=&\frac{\beta J}{N}, \\
\tilde{\Delta}&=&\frac{\beta \Delta}{N}+\frac{1}{2}\log \left [\frac{1}{2}\sinh\left (\frac{2\beta \Omega}{N}\right)\right ],
\end{eqnarray}
where $N$ is the Trotter number and
we adopt the periodic boundary condition for the Trotter direction $\sigma_{N+1}=\sigma_1$.

We consider the $\mu$-th chain along the Trotter direction whose Hamilton is given by
\begin{eqnarray}
\beta {\cal H}_{\rm eff}^{\mu}=-K_{N}\sum_{i=1}^{N}\sigma_{i,\mu}\sigma_{i+1,\mu}
-\sum_{i=1}^NH_i\sigma_{i,\mu}+\tilde{\Delta}\sum_{i=1}^N\sigma_{i,\mu}^2,
\end{eqnarray}
where $H_i=\tilde{J}\sum_{\langle \mu\nu\rangle}\sigma_{i,\mu}$. 
We define the following transfer matrix:
\begin{eqnarray}
{\cal T}_n(\sigma_N,\sigma)&=&\sum_{\sigma_1,\sigma_2,\cdots,\sigma_{n-1}}
\exp\{K_{N}(\sigma_N\sigma_1+\sigma_1\sigma_2+\sigma_2\sigma_3+\cdots+\sigma_{n-1}\sigma)\nonumber \\
&&+(H_1\sigma_1+H_2\sigma_2+\cdots +H_n\sigma)\nonumber \\
&&-\tilde{\Delta}(\sigma_1^2+\sigma_2^2+\cdots +\sigma^2)\}.
\end{eqnarray}
${\cal T}_n(\sigma_N,\sigma)$ can be readily obtained from the recursion formula
\begin{eqnarray}
{\cal T}_n(\sigma_N,\sigma)=\sum_{\sigma_{n-1}}{\cal T}_{n-1}(\sigma_{n-1},\sigma)\exp\{K\sigma_{n-1}\sigma+H_n\sigma-\tilde{\Delta}\sigma^2\}, (n\ge2)
\end{eqnarray}
with ${\cal T}_1( \sigma_N,\sigma)=\exp (K_N\sigma_N\sigma+H_1\sigma_1-\tilde{\Delta}\sigma^2)$.
The probability $P_N(\sigma_N=0, \pm 1)$ that the boundary spin $\sigma_N$ takes $0, \pm 1$ is given as
\begin{eqnarray}
P_N(\sigma_N)=\frac{{\cal T}_N(\sigma_N,\sigma_N)}{\sum_{\sigma}{\cal T}_N(\sigma,\sigma)}.
\end{eqnarray}
We can determine the value of $\sigma_N$ by using this probability.
The other spins $\sigma_{k} (k=N-1, N-2, \cdots, 1)$ are determined by the probability given by
\begin{eqnarray}
P_{k}(\sigma_{k})=\frac{\tilde{\cal T}_{k}(\sigma_N,\sigma_{k})}{\sum_{\sigma}\tilde{\cal T}_N(\sigma_N,\sigma)},
\end{eqnarray}
where
\begin{eqnarray}
\tilde{\cal T}_{k}(\sigma_N,\sigma)={\cal T}_{k}\exp(K\sigma\sigma_{k+1}).
\end{eqnarray}
Thus, we determine new spin configuration of the chain.

\begin{figure}
\epsfile{file=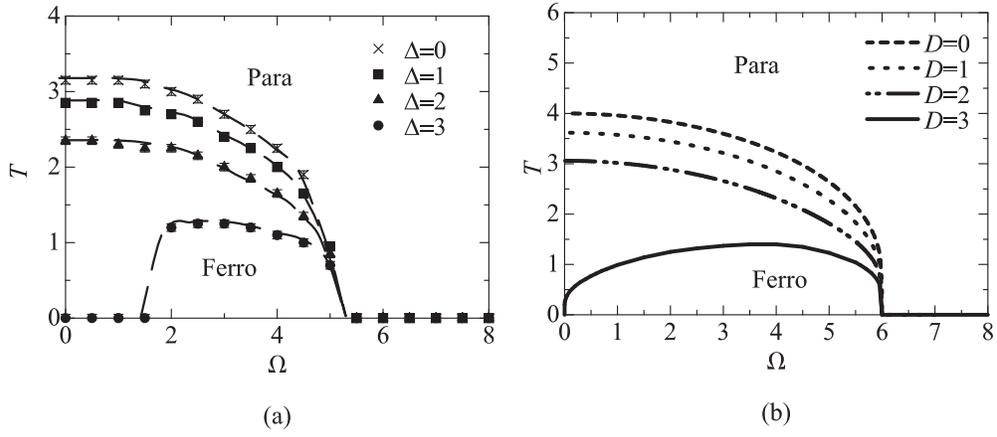,width=14cm}
\caption[]{Phase diagram in the cordinate ($\Omega, T$) by using QMC simulation (a) and MFA (b) \label{fig8}}
\end{figure}

In Fig. \ref{fig8}(a) we present the phase diagram in the coordinate $(\Omega, T)$ obtained by the QMC simulation.
In Fig. \ref{fig8}(b) the phase diagram obtained by MFA is also given,
We confirmed that the qualitative feature of the phase diagrams is common in these two cases.
Here it is to notify that a ferromagnetic phase appears by being induced by quantum fluctuation in the case of large $\Delta$ region ($\Delta\simeq 3$).
That is, at $\Delta=3$ the ferromagnetic phase is suppressed in classical BC model ($\Omega=0$), but it appears at finite value of $\Omega$.
\end{appendix}

\section*{ACKNOWLEDGMENTS}
The authors would like to thank Prof. Y. Noda of Tohoku Univ. for allowing us to use their experimental data prior to publication.

The numerical calculations 
were performed on the supercomputers in
computational facility of the Super Computer 
Center of Institute for Solid State Physics,
University of Tokyo.

\begin{table}
\caption{}{\label{table1}}
\begin{tabular}[h]{c|c|c}
        &  quantum displacive  & quantum order-disorder \\ \hline
  $T_0$ & $\frac{1}{k_{\rm B}}\left ( \frac{|\omega_0^2|\omega_{\rm loc}^2}{\lambda}\right )$ & $\frac{J(0)}{k_{\rm B}}$\\ \hline
  $T_1$ & $\frac{\hbar\omega_{\rm loc}}{k_{\rm B}}$ & $\frac{2\hbar\Omega}{k_{\rm B}}$ \\ \hline
  $C$   & $\frac{\omega_{\rm loc}^2}{k_{\rm B}\lambda}$ & $\frac{1}{k_{\rm B}}$\\
\end{tabular}
\end{table}

\newpage
\newpage
\newpage
\newpage
\newpage
\newpage

\newpage

\end{document}